\documentclass[twocolumn,prl,aps,superscriptaddress]{revtex4-1}
\usepackage{amsmath, amsthm, amssymb}
\usepackage{pifont}
\usepackage{dsfont}
\usepackage{amscd}
\usepackage{amsfonts}
\usepackage{textcomp}

\usepackage{float}
\usepackage{mathpazo}

\fontfamily{palatino-ttf}

\usepackage[pdftex]{graphicx}
\usepackage[usenames,dvipsnames]{color}

\usepackage{mathpazo}
\newcommand{\be}{\begin{equation}}
\newcommand{\ee}{\end{equation}}
\newcommand{\ben}{\begin{enumerate}}
\newcommand{\een}{\end{enumerate}}
\newcommand{\defin}{{\bf{Definition: }}}
\newcommand{\half}{\frac{1}{2}}
\newcommand{\chn}{{\cal N}}
\newcommand{\coh}{I_c(A\rangle B)}

\newcommand{\pdit}{{\gamma^{(d)}}}

\newtheorem{lemma}{Lemma}

\fontfamily{palatino-ttf}

\begin{document}

\title{When does noise increase the quantum capacity?}
\author{Fernando G.S.L. Brand\~ao}
\affiliation{Departamento de F\'isica, Universidade Federal de Minas Gerais, Belo Horizonte, Caixa Postal 702, 30123-970, MG, Brazil}
\author{Jonathan Oppenheim}
\affiliation{Department of Applied Mathematics and Theoretical Physics, University of Cambridge, Cambridge CB3 0WA, U.K.}
\author{Sergii Strelchuk}
\affiliation{Department of Applied Mathematics and Theoretical Physics, University of Cambridge, Cambridge CB3 0WA, U.K.}

\begin{abstract}
Superactivation is the property that two channels with zero 
quantum capacity can
be used together to yield positive capacity.  Here we demonstrate
that this effect exists for a wide class of inequivalent channels, none of which
can simulate each other.
We also consider the case where one of two zero capacity channels is applied, but the sender is ignorant of which one is applied. We find examples where the greater the entropy of mixing of the channels, 
the greater the lower bound for the capacity. Finally, we show that the effect of superactivation is rather generic by providing an example of superactivation using
the depolarizing channel.
\end{abstract}
\maketitle

A quantum channel is any physical process which can be applied to a quantum system.
There is an input to the channel, and we are interested in how much
information remains at the output. Some channels are so noisy that no quantum information can be reliably transmitted
through them -- error correction becomes impossible and one cannot send a quantum state through the channel faithfully. 
We say that such channels
have {\it zero capacity}. In classical information theory, zero capacity channels
are not interesting, because they only include the case where there is no correlation
between the input and output. However, some zero capacity quantum channels have 
surprising properties: for example, they can be used to share a private key~\cite{pptkey,horodecki_qkdprivst},
and two zero capacity channels can be combined in parallel to reliably send quantum states, a
situation that is impossible classically~\cite{smith_superact}.

The ability to send quantum information down two channels which have zero capacity is called
{\it{superactivation}}, and it is an important phenomenon which suggests that quantum channels are radically different from classical ones.
For classical channels, we can quantify a channel by its capacity, while the phenomena of superactivation
means that for a quantum channel, the capacity does not adequetly characterize the channel, since
the utility of the channel  depends on what other channels are also available.  One hopes that a greater understanding
of superactivation will enable progress to be made in understanding the quantum capacity, something made difficult because
we still do not have an adequate formula for it.  Additionally there appear to be strong
links between superactivation and privacy~\cite{smith_extensive_2009}, and these are not yet properly understood.

Despite the importance of superactivation,  only one example is known~\cite{smith_superact}: one of the channels is a symmetric
channel, meaning that the quantum state of the output and the environment is symmetric under exchange. 
This channel cannot be used for quantum communication because its symmetry implies that if this channel had positive quantum capacity it would violate the no-cloning theorem~\cite{WoottersZ-cloning}. An example
is the $50\%$ erasure channel, denoted as ${\cal N}_e^{0.5}$, which faithfully transmits the input state half of the time and outputs an erasure flag in the rest of the cases.  The only known protocol for superactivation 
involved using the $50\%$ erasure channel.  The second channel is one which produces a private
key, but cannot be used to send quantum information~\cite{pptkey}.  Such a channel is known
to have zero-capacity because it has a positive partial transpose (PPT)~\cite{peres_separabilitycr}, which
implies that it has zero capacity~\cite{horodecki_bec}.

It was also shown in~\cite{smith_superact} that a convex combination of ``flagged'' channels 
\be\label{convcomb}
{\cal N} = \kappa {\cal N}_{{\gamma}^{(d)}} \otimes |0\rangle \langle 0|_B + (1-\kappa){\cal N}_e^{0.5} \otimes |1\rangle \langle 1|_B,
\ee
has positive quantum capacity for a particular private channel ${\cal N}_{{\gamma}^{(d)}} $, and for a very small amount of mixing ($\kappa=0.0041$). 

It is natural to ask about the generality of this phenomenon. First, whether there exist communication protocols that allow for strong nonconvexity of quantum capacity, in the sense that $\kappa$ can have a large range. Indeed, we will
find here that one can achieve positive capacity for any $0<\kappa<1$.  This surprising result 
implies that a generic mixing of the zero capacity channels during the transmission will, nevertheless, increase the quantum capacity.  In fact
we find situations where, counter-intuitively, the more noise, the greater is the lower bound for the capacity given by the so-called \textit{coherent information}.
A second question we address is what types of
channels can be superactivated.  Since there are very limited techniques to show a channel has zero capacity, this
is a difficult problem. 
It was not presently known whether this startling effect can be generalised to any channels other than ${\cal N}_e^{0.5}$. 
Here we find that superactivation is possible for a large class of inequivalent and generic channels (in the sense that they cannot
simulate each other).  This includes erasure channels with any probability $p\in [\half, 1)$ of erasure, as well as the common depolarizing 
channel ~\cite{depolarizing}.
Third, 
we are interested in whether superactivation is robust against noise or can only be demonstrated using perfectly noiseless resources.  This is particularly important in lieu of proposed experiments to test this effect~\cite{smith_gaussian_2011}. 
We answer this question affirmatively.

It is of course a basic question in quantum information theory to quantify the ability of quantum channels to transmit quantum states faithfully. The former is described mathematically as a completely positive trace preserving map ${\cal N}: A\to B$ from density matrices on input system $A$ to density matrices on an output system $B$. The performance of a quantum channel for noiseless quantum communication is characterised by its \textit{quantum capacity} $\cal Q(N)$, which is the maximum achievable rate for quantum communication. Analogously, $\cal Q(N)$ quantifies the amount of pure state entanglement that can be transmitted through ${\cal N}$.

The quantum capacity is known to be lower bounded by the \textit{coherent information}~\cite{Lloyd-cap,shor-cap,devetak_privatecc}:
\be
{\cal Q(N)}\ge  I_c(A\rangle B) :=\displaystyle\max_\rho (S(B)_\sigma -S(E)_\sigma),
\ee
where the von Neumann entropies are evaluated on $\sigma_{BE} = U \rho U^\dagger$, with $U : A \mapsto BE$ the isometry associated to the channel ${\cal N}$ as follows: ${\cal N}(\rho) = \text{tr}_{E}(U \rho U^{\cal y})$.
The first family of zero-capacity channels we will consider, denoted as ${\cal N}_{\gamma^{(d)}}$, produce bound entangled states -- states that need pure state entanglement to create them, but from which no pure state entanglement can be
extracted~\cite{horodecki_BE}. Such states, despite being useless for transmission of quantum information, may contain secrecy~\cite{pptkey}. Here we take  ${\cal N}_{\gamma^{(d)}}$ to be such a channel which produces bound entangled states
that contain secrecy and in particular ``private bits''.

{\bf{Private bits and coherent information.}} Quantum states that contain $d$ bits of secrecy are called private dits, pdits, or twisted ebits~\cite{pptkey,horodecki_general_2009} and have the generic form
\be\label{pdit}
\gamma^{(d)}=UP_{AB}^+\otimes \sigma_{A^{'}B^{'}}U^\dagger,
\ee
where $U= \sum_{i,j=0}^{d-1} |ii\rangle\langle jj|_{AB} \otimes U_{ij}$ is a controlled unitary operation termed {\textit{twisting}} (with arbitrary unitaries 
$U_{ij}$), $P^{+}_{AB}$ is the projector onto a $d$ dimensional maximally entangled state, and $\sigma_{A^{'}B^{'}}$ is an arbitrary state called the ``shield'' subsystem of dimension $d'$, for its presence protects private correlations. In the case when $d=2$ we will call it a pbit. Parties that 
have $A$ and $B$ subsystems of a pdit (known as the ``key'') 
can extract $\log_2d$ ebits by performing $U^\dagger$ if one of them possesses the shield $A'B'$ in its entirety. However, when
the shield is split between the two parties, it can be impossible to perform the untwisting using only local operations, and there exist
states which are arbitrarily close to pdits, yet no ebits can be produced from them.
The main idea we will be exploiting here
is that superactivation can occur by one zero capacity channel being used to share pdits, and then by Alice using a second zero-capacity channel to send her part of the shield $A$ to Bob some of the time so that he can perform the untwisting operation, giving them shared ebits~\cite{O08-superactivation} on these occasions.

We will thus consider using  ${\cal N}_{\gamma^{(d)}}$ in conjunction with a number of different channels: first, erasure channels ${\cal N}_e^{p}$,
which outputs an erasure flag with probability $p \in [\frac{1}{2},1)$,  and faithfully transmits the
input state otherwise.  These are all inequivalent channels, in the sense that for $p\in \{1-\frac{1}{n}| n\in \mathbb{N} \mbox{\textbackslash}\{1\}\}$ no such channel
with probability $p$ can simulate one with probability of erasure smaller than $p$~\cite{hastings_infinitely_many_chn}.
Moreover, it is known that ${\cal N}_e^p$ retains zero capacity in this range since a higher erasure
probability can only decrease the capacity. Our results hold for all $p \in [\frac{1}{2},1)$.

{\bf{Strong nonconvexity of quantum capacity.}} Consider the convex combination
of two channels
as in Eqn. \ref{convcomb}, where $\chn_\pdit$ is the PPT channel that generates
noisy pdits, which can be made arbitrarily close to perfect pdits at the expense
of increasing the dimension of the shield, and the erasure probability of the
latter channel is in the range $p \in [\frac{1}{2},1)$. We take the input dimension of both
channels to be equal.  For clarity of presentation, we will consider the
limiting case, when the dimension of the shield goes to infinity, and take the
key part to be perfect.
Both the PPT pdit channel and the erasure channel have zero quantum capacity.
The quantum capacity of the resulting mixture of the two channels can be
strictly positive when $p=0.5$, and $\kappa \in (0;0.0041)$~\cite{smith_superact}. We 
now show that this is much more generic, and
will employ the protocol described below to show that
for the PPT pdit channel and $50\%$ erasure channel in the convex mixture we
can surprisingly achieve positive quantum capacity for all $\kappa\in(0,1)$.

More formally, consider a channel $\cal N$ in the form of Eqn. (\ref{convcomb}) and consider the following protocol:
\ben
\item Alice initially feeds $d+d'$ halves of ebits through $\cal N$, keeping 
the other halves of the ebits -- for clarity, we denote the subsystems which are kept 
in Alice's possession as $AA'$ with $A$ of dimension $d$ and $A'$ of dimension $d'$.  The 
subsystems at Bob's side after the
transmission will be $BB'$. 
If this is repeated $n$ times, then at the end of this step they share $n$ instances of $AA'BB'$.
\item Alice feeds her instances of $A'$ into the channel, and pads her input with $d$ fresh qubits which will 
not play any role in this round of the protocol and are discarded by Bob. After
the transmission Alice and Bob have $n$ instances of subsystems $A$ and
$BB'\tilde{A'}$ respectively.
\een
At the end of the protocol we get:
\begin{equation}\label{nonconvexp}
I_c(A\rangle B)_{\gamma^{(d)}} = \half(1-\kappa)\left[\kappa-p(\kappa+2)+1\right]\log d.
\end{equation}
When $p=0.5$ the expression for the coherent information simplifies to:
\begin{equation}
I_c(A\rangle B)_{\gamma^{(d)}}=\frac{1}{4}(1-\kappa)\kappa \log d.
\end{equation}
See Appendix A in Supplemental Material for the calculation of the coherent information. 
\begin{figure}[h]
\includegraphics[scale=0.64]{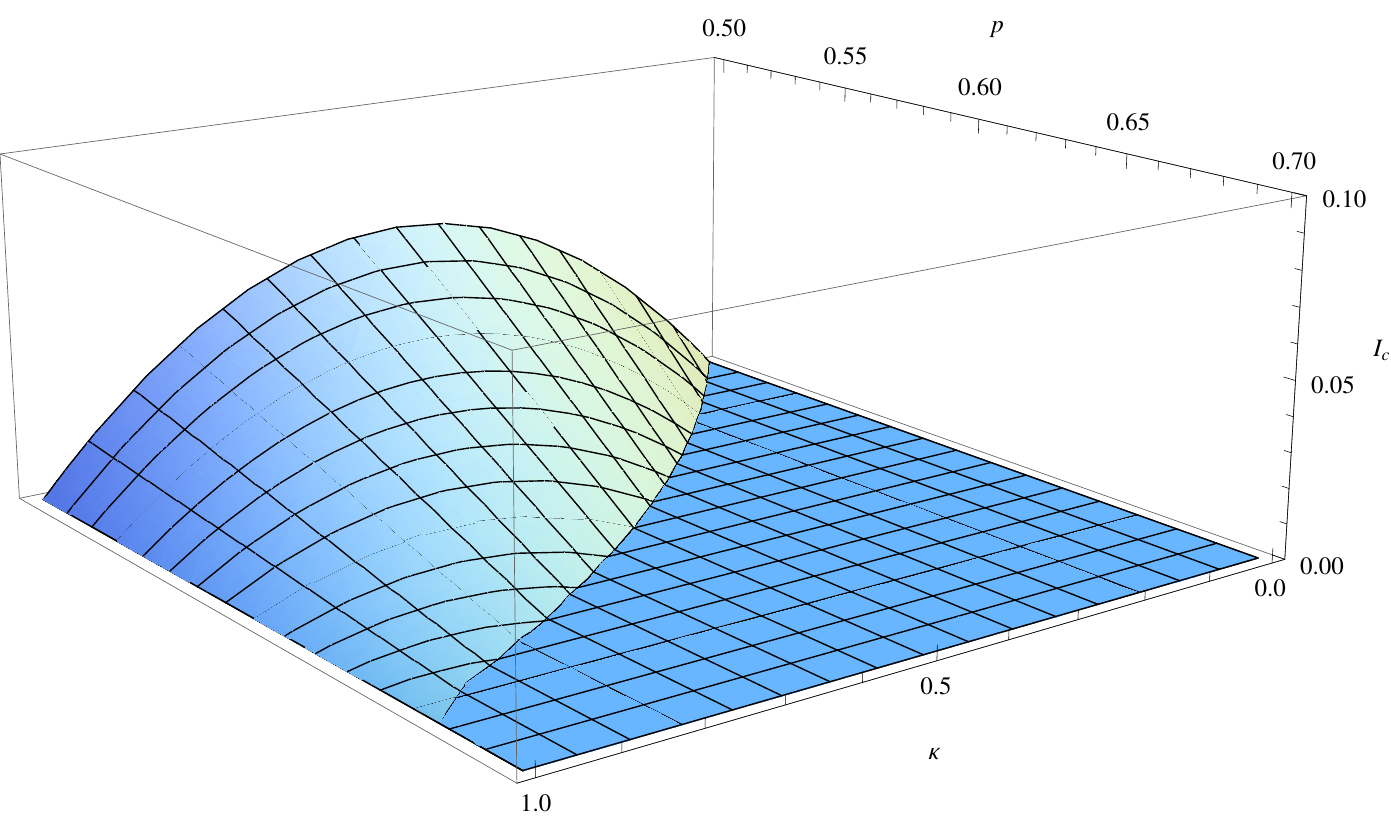}
\caption{Nonconvexity of quantum capacity for $I_c(A\rangle B) =
\half(1-\kappa)\left[\kappa -p(\kappa+2)+1\right]\log d$ when $d=2$ when the dimension of the shield subsystem tends to infinity.}
\label{mainplot}
\end{figure}
Fig. \ref{mainplot} demonstrates the full range of pairs $(\kappa, p)$ for
which the violation of the convexity of quantum capacity is achieved.

The full nonconvexity of the coherent information for the convex combination
\eqref{convcomb} holds when $p=0.5$, when the dimension of the shield subsystem tends to infinity, and is not true for larger $p$. 
This is also where the greater the entropy of mixing of the two channels, the greater the lower bound for the capacity given by the coherent information.

{\bf{Inequivalent classes of superactivating channels with noisy resources.}} We next address 
the question of generalizing the superactivation example to the class of erasure channels with $p>\half$, and we will simultaneously 
tackle the question of robustness of superactivation to noise.  We do so by establishing the region of pairs $(p,\epsilon)$, where $p \in [\frac{1}{2},1)$ is the erasure probability, and $\epsilon$ denotes the amount of tolerable noise in the PPT pbit channel, for which we can demonstrate superactivation.

In the two-step protocol~\cite{O08-superactivation} that achieves superactivation for an arbitrary pbit channel ${\cal N}_{\gamma^{(d)}}$,
Alice and Bob first use the pbit channel  to share states of Eqn. 
\eqref{pdit}, then in the second step Alice sends her part of the shield (subsystem $A^{'}$) through the erasure channel ${\cal N}_e^{0.5}$. 
Half of the time, when the erasure does not take place, Bob is able to perform the $U^\dagger$ of Eqn. \eqref{pdit} and they end up sharing an ebit. When erasure occurs, they are left with
a classically correlated state and an erasure flag.  We now show that this protocol works for other values of $p$.
Since the case of erasure (non-erasure) is distinguishable on Bob's site, the 
lower bound for the capacity of the joint channel ${\cal N}_\gamma\otimes {\cal N}_e^p$ is just the coherent information averaged over the two
cases: 
\be\label{plowerbound}
{\cal Q}({\cal N}_{\gamma^{(d)}}\otimes {\cal N}_e^{p}) \ge pI_{c}(A\rangle B)_{\gamma^{(d)}_{er}}+(1-p)I_{c}(A\rangle B)_{\gamma^{(d)}_{uner}},
\ee
where $p=0.5$ in the original example~\cite{smith_superact}, and the first term is evaluated on the state ${\gamma^{(d)}_{er}}$ that corresponds to the case when Bob received the erasure flag while the latter is evaluated on ${\gamma^{(d)}_{uner}}$, when Alice's share of the shield was successfully transmitted to Bob. If the erasure event takes place, and the shield doesn't get through, Bob will not be able to undo the unitary $U$, so $I_{c}(A\rangle B)_{\gamma^{(d)}_{er}}=0$. If shield gets through, assuming operations are perfect, $I_{c}(A\rangle B)_{\gamma^{(d)}_{uner}} = \log d$. In the case of many copies, Alice and Bob will share $m=(1-p)n$ pdits on average and 

\be\label{icohall}
I_{c}(A\rangle B)_{\left(\gamma^{(d)}\right)^{\otimes m}} =mI_{c}(A\rangle B)_{\gamma^{(d)}_{uner}}=(1-p)n\log d.
\ee
This is under the assumption that the pbits are perfectly private, and so to investigate what happens when this restriction is lifted, 
we consider channels which produce approximate pbits:\\
\defin The state $\widetilde{\gamma}^{(d)}$ is called an {\textit{$\epsilon$-approximate pdit}} if there exists a set of measurement operators on the key subsystem $\{P_i^{AB}\}_{i=1}^d$ such that 
\be\label{noisydef}
\left|\mbox{Tr}_{A^{'}B^{'}}\left(\sum_{i}P_i \widetilde{\gamma}^{(d)} P_i\right)-K_{AB} \otimes M_{E}\right| \le \epsilon,
\ee where $K_{AB}$ represents the key subsystem and $M_E$ represents the environment.\\
An approximate pdit satisfies the following property:
For every $\widetilde{\gamma}^{(d)}$ there exists a unitary $U = \{U_{ij}\}$ on the system such that \begin{equation}
\left |U^\dagger \widetilde{\gamma}^{(d)} U - \Psi^+_{AB}\otimes \sigma_{A^{'}B^{'}}\right| \le \epsilon
\end{equation}
This follows directly from Theorem 2 in~\cite{horodecki_general_2009}. From now on, we will limit the set of all approximate pdits to the subset of the approximate pdits which have PPT. The existence of good PPT approximations of pdits is shown in
\cite{pptkey}.
\\
Following the same protocol as in~\cite{smith_superact,O08-superactivation}, consider a pair of channels \be\label{noisysup}
{\widetilde{\cal N}}_{\gamma^{(d)}}\otimes {\cal N}_e^p,
\ee
with $p \in [\frac{1}{2},1)$, where using ${\cal \widetilde{N}}_{\gamma^{(d)}}$ results in Alice and Bob sharing an $\epsilon$-approximate pdit $\widetilde{\gamma}^{(d)}$.
Then Alice sends her share of the shield to Bob using ${\cal N}_e^p$ as above. After many independent uses of ${\widetilde{\cal N}}_{\gamma^{(d)}}$ they share $m=(1-p)n$ copies of $\widetilde{\gamma}^{(d)}$. 
The question of interest is whether given a large number $n$ of $\widetilde{\gamma}^{(d)}$ Alice and Bob could superactivate them
with an erasure channel of probability $p$, i.e. whether there exist pairs $(p, \epsilon)$ which will make the lower bound 
on the quantum capacity given by Eqn. \eqref{noisysup} strictly positive. The following lemma will make use of Eqn. \eqref{plowerbound} and relation \eqref{noisydef} to derive a lower bound on the joint channel of Eqn. \eqref{noisysup}:
\begin{lemma}\label{mainlemma} Consider independent uses of ${\cal \widetilde{N}}_{\gamma^{(d)}}\otimes {\cal N}_e^p$, $p \in [\frac{1}{2},1)$. 
Then 
\be {\cal Q}({\cal \widetilde{N}}_{\gamma^{(d)}}\otimes {\cal N}_e^p) \ge (1-p-4\epsilon)\log d-2h(\epsilon),\ee where $d$ is the dimension of the key part, and $h(\cdot)$ is a binary entropy.\end{lemma}
 See Appendix B in Supplemental Material for the proof and graphical illustration.

{\bf{Superactivation using depolarizing channel.}} It turns out that the erasure channel and its variants are not the only channels that can be used in conjunction with the PPT pbit channel for superactivation.  Here we consider also ${\cal \widetilde{N}}_{\gamma^{(d)}}\otimes\chn_{dep}$, with $\chn_{dep}$ the commonplace depolarizing channel~\cite{depolarizing} given by

\be\label{depo}
\chn_{dep} = p\chn_{id} + (1-p)\chn_{mix}.
\ee
The first channel in this mixture is the identity channel acting as $\chn_{id}(\rho)=\rho$, and the second one is the completely randomizing channel acting as $\chn_{mix}(\rho)=\frac{\mathbb{1}}{r}$. The depolarizing channel is so ubiquitous in part because all quantum channels can be twirled to this form by applying some randomly chosen bilateral unitary to the input and output of the channel ~\cite{depolarizing}. It follows that $\chn_{dep}$, for arbitrary input dimension $r$, is anti-degradable and thus has zero capacity in the range $p\in [0;\frac{1}{2}]$. This follows from the fact that the Jamiolkowski state associated to the channel $1/2(P_{AB}^{+} + \mathbb{1}_{AB}/r^{2})$ has a two-symmetric-extension, namely $1/2 (P_{AB}^{+} \otimes \mathbb{1_{B'}}/r  + P_{AB'}^{+} \otimes \mathbb{1_{B}}/r)$. Remarkably, we will
find that this channel can be used for superactivation, even as the amount of noise is made arbitrarily large.

The superactivation protocol is as before - after creating approximate pbits using ${\cal \widetilde{N}}_{\gamma^{(d)}}$, 
Alice sends the shield $A^{'}$ to Bob through the depolarizing channel.
Unlike the previous examples of erasure channels,
there are no flags attached to the output, so Bob doesn't know which channel was applied.
After the transmission, Alice and Bob are left with the mixture of two states: with probability $p$, after Bob performing the untwisting operation 
$U^\dagger$, 
they share the maximally entangled state $\Phi^{+}_{AB}$, and with probability $(1-p)$ the ebits cannot be untwisted and 
they share the an approximation $\sigma_{AB, \epsilon}$ classically correlated state $\sigma_{AB} := 1/d \sum_k |k, k\rangle \langle k, k|$, i.e. they share the state
\be
\omega_{AB} =p\Phi^{+}_{AB} + (1-p) \sigma_{AB, \epsilon}.
\ee
The fact that we only get an approximation $\sigma_{AB, \epsilon}$ of the classically correlated state is due to the fact that the channel 
$\chn_{\gamma^{(d)}}\otimes\chn_{dep}$ only created approximate pbits. For any $\epsilon > 0$ we can choose the dimension of the shield state and of the 
depolarizing channel sufficiently large so that $\Vert \sigma_{AB, \epsilon} - \sigma_{AB} \Vert_1 \leq \epsilon$. The coherent information, evaluated on $\omega_{AB}$ for $d = 2$, can be lower bounded as follows
\begin{align}\label{superdepo}
\coh_{\omega_{AB}} \geq & 1 +\frac{1-p}{2}\log\left(\frac{1-p}{2}\right) \\&+\frac{1+p}{2}\log\left(\frac{1+p}{2}\right) - 4\epsilon \log(d)+2h(\epsilon).
\end{align}
This follows by computing the coherent information for $p\Phi^{+}_{AB} + (1-p) \sigma_{AB}$ and using Fannes inequality and the relation $\Vert \sigma_{AB, \epsilon} - \sigma_{AB} \Vert_1 \leq \epsilon$. For any fixed $p$ we can take the dimension of the depolarizing channel and of the shield part of $\chn_{\gamma^{(2)}}$ sufficiently large so that $\epsilon$ is as small as we wish. 
In this regime we find superactivation for a large region of values of $p$ in the range $(0, \half]$, which constitute new examples of superactivation using the depolarizing channel (see Appendix C in Supplemental Material for the plot of the region for $(p,\epsilon(p))$).

We have seen that superactivation doesn't only occur for the two special channels considered in the initial discovery of the effect.  Rather,
there are classes of generic and common channels, as well as inequivalent ones which can be used for superactivation and, likewise, for the curious effect where adding noise (by increasing the entropy of mixing of two channels) can increase the quantum capacity.  Here too, we find that it is not
a tiny mixture of noise which increases the capacity, but rather, there are cases where the more the noise, the greater the capacity, and generally
any amount of mixing can result in positive capacity.   Although we have found superactivation to be more generic than previously thought,
we have only considered cases where one channel has zero capacity because it is PPT, and the other channel has zero capacity because of the
no-cloning bound.  The big question of whether superactivation exists for channels which don't each belong to these classes rests unanswered.
This is a challenging question since at the moment we have no other ways of showing a channel has zero capacity.  We hope the considerations
here provide some clues to the answer.
\newpage
\begin{center}
\Large{\textbf{Supplemental Material}}
\end{center}

{\textbf{Appendix A: Calculation of The Coherent Information in the Strong Nonconvexity Protocol.}}

The action of the convex combination can be represented as the collection of
three channels acting on the input state with certain probabilities: the
private channel $\chn_\pdit$, the identity channel ${\cal I}$ (when the input
state goes through the erasure channel and erasure does not occur), and the
erasing channel ${\cal E}$ (when the input state goes through the erasure
channel and erasure occurs deterministically).
The coherent information for these 
9 possible situations, is presented in the table below:
\begin{table}[H]\caption{Channels acted and the corresponding coherent
information at the end of the protocol}
\label{T1}
\centering
\begin{tabular}{ |c|c| }
\hline Coherent information & Channels acted  \\
\hline $\kappa(1-\kappa)(1-p) \log d$ & ${\cal N}_{\gamma^{(d)}}\otimes {\cal
I}$  \\
\hline $\kappa(1-\kappa)(1-p) \log d$ & $ {\cal I}\otimes {\cal
N}_{\gamma^{(d)}}$  \\
\hline $0$                                 & $ {\cal N}_{\gamma^{(d)}}\otimes
{\cal N}_{\gamma^{(d)}}$\\
\hline $(1-\kappa)^2(1-p)^2 \log d$   & ${\cal I} \otimes {\cal I}$ \\
\hline $(1-\kappa)^2 p(1-p) \log d$        & ${\cal E} \otimes {\cal I}$ \\
\hline $-(1-\kappa)^2p(1-p) \log d$        & ${\cal I} \otimes {\cal E}$ \\
\hline $-(1-\kappa)^2p^2 \log d$        & ${\cal E} \otimes {\cal E}$ \\
\hline $-\kappa(1-\kappa)p\log d$      & ${\cal N}_{\gamma^{(d)}}\otimes {\cal
E}$\\
\hline $0$      & ${\cal E}\otimes {\cal N}_{\gamma^{(d)}}$\\
\hline  \end{tabular}\end{table}

Since which channel acted is flagged on Bob's site, as well as 
whether the input was erased in the case where the erasure channel acted, the coherent information
is just a sum of all the entries of the table, divided by $2$ since the protocol takes $2n$ uses of the channel.\\

{\textbf{Appendix B: Superactivation with the Large Class of Erasure Channels.}}

{\textbf{Proof of Lemma 1:}}
We will first examine the case when the shield was successfully transmitted to Bob (erasure does not take place). 
This case amounts to evaluating $I_{c}(A\rangle B)_{\widetilde\gamma^{(d)}_{uner}}$. 
The application of the Alicki-Fannes inequality for quantum conditional entropy~\cite{alicki_fannes} gives us:
\begin{align}
|&I_{c}(A\rangle B)_{\gamma^{(d)}_{uner}}-I_{c}(A\rangle B)_{\widetilde\gamma^{(d)}_{uner}}| \\
&=|S(A|B)_{\gamma^{(d)}_{uner}}-S(A|B)_{\widetilde\gamma^{(d)}_{uner}}|\\
&\le 4\epsilon\log d+2h(\epsilon).
\end{align}\\\\
Using Eqn. (7) in the Letter: $I_{c}(A\rangle B)_{\gamma^{(d)}_{uner}}=\log d$, gives
\begin{align}
 I_{c}(A\rangle B)_{\widetilde\gamma^{(d)}_{uner}}\ge(1-4\epsilon) \log d-2h(\epsilon).
\end{align} 
When erasure takes place, we similarly obtain the lower bound for $I_{c}(A\rangle B)_{{\widetilde\gamma}^{(d)}_{er}}$ by noting that in this case $I_c(A\rangle B)_{\gamma^{(d)}_{er}}=0$, thus: \be
 I_{c}(A\rangle B)_{{\widetilde\gamma}^{(d)}_{er}}\ge-4\epsilon \log d-2h(\epsilon).
\ee
Recalling that, analogously to Eqn. (6) in the Letter, we have $I_c(A\rangle B)_{{\widetilde\gamma}^{(d)}} = (1-p)I_c(A\rangle B)_{{\widetilde\gamma}^{(d)}_{uner}}+ pI_c(A\rangle B)_{{\widetilde\gamma}^{(d)}_{er}}$, and substituting the expressions for the coherent information we obtain the final result: 
\begin{align}
{\cal Q}({\cal \widetilde{N}}_{\gamma^{(d)}}\otimes {\cal N}_e^p) &\ge I_c(A\rangle B)_{{\widetilde\gamma}^{(d)}} \\
&\ge (1-p-4\epsilon)\log d-2h(\epsilon).
\end{align}
 $\blacksquare$

\begin{figure}
\centering
\includegraphics[scale=0.65]{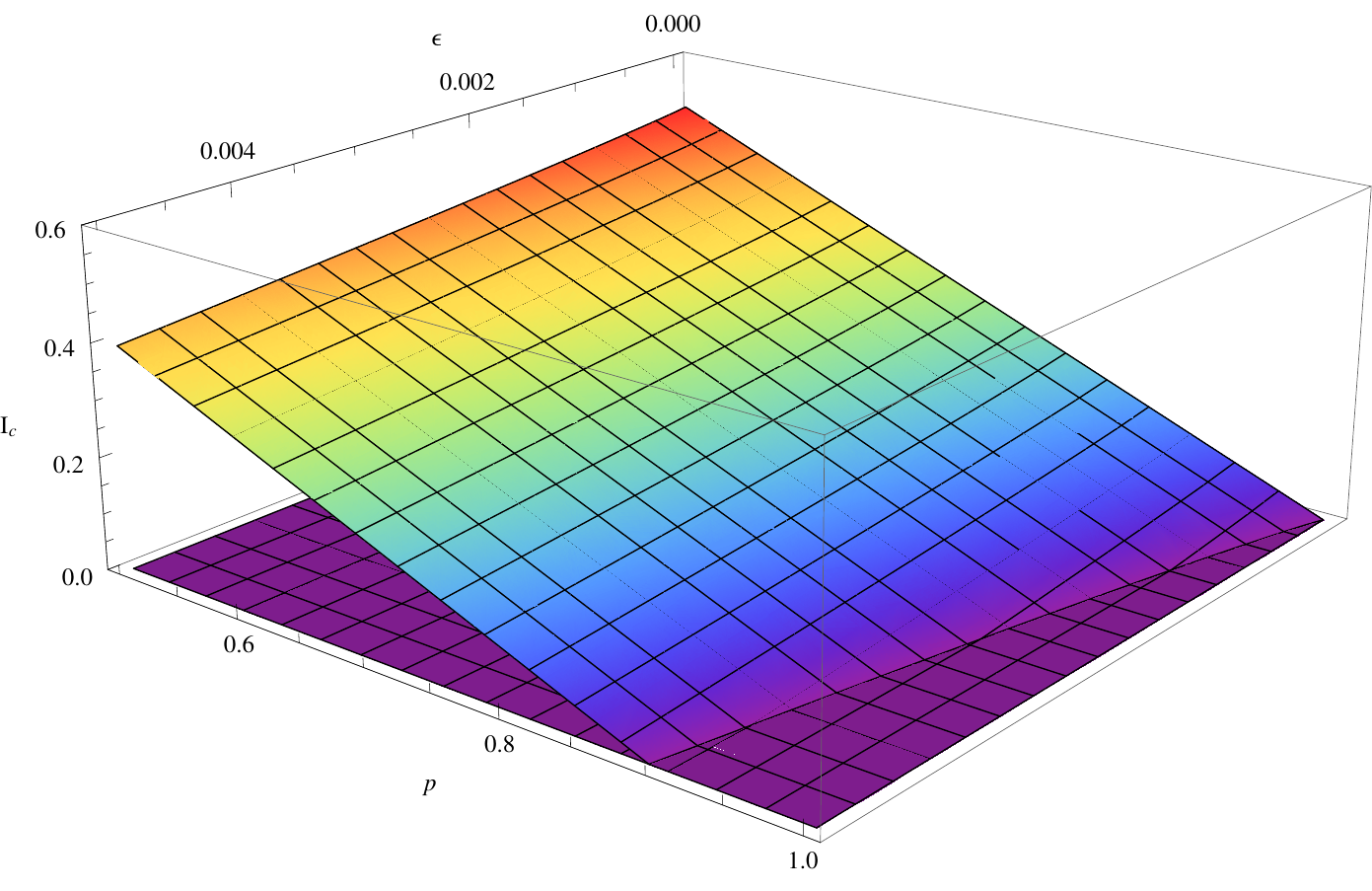}

\caption{The lower bound on quantum capacity $g(p,\epsilon)= (1-p-4\epsilon)\log d-2h(\epsilon)$ for $p\in [\frac{1}{2},1)$ and $\epsilon=\epsilon(p).$ }
\label{noisysa}
\end{figure}
As Fig. \ref{noisysa} demonstrates, for each channel ${\cal N}_e^p$ with the erasure probability $p\in [\frac{1}{2},1)$ there exists $\epsilon = \epsilon(p)$, and the pair $(p, \epsilon(p))$ satisfies $I_c(A\rangle B)_{{\widetilde\gamma}^{(d)}}>0$.

{\textbf{Appendix C: Superactivation With Depolarizing Channel.}}

\begin{figure}[h]
\centering
\includegraphics[scale=0.65]{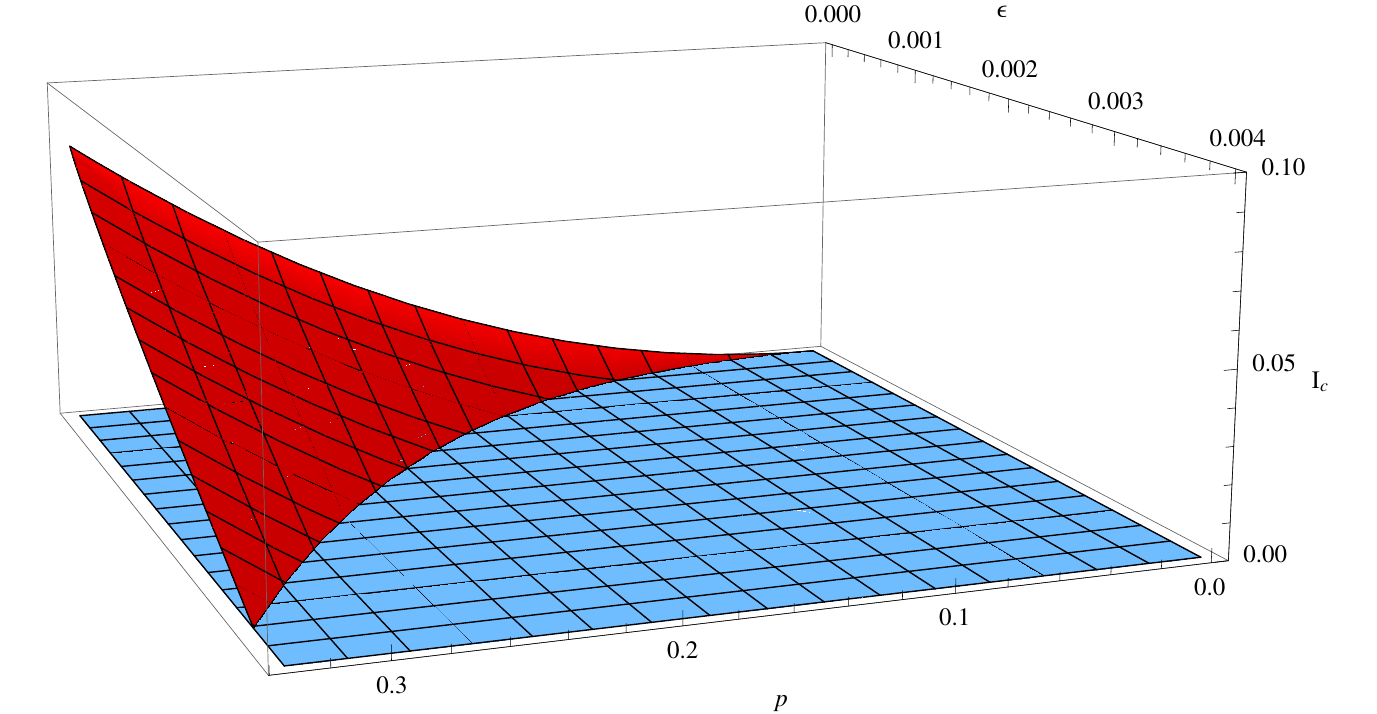}
\caption{Superactivation using depolarizing channel with noisy resources.}
\label{sup_depol_with_noise}
\end{figure}

\bibliographystyle{apsrev}

\begin{thebibliography}{18}%
\makeatletter
\providecommand \@ifxundefined [1]{%
 \@ifx{#1\undefined}
}%
\providecommand \@ifnum [1]{%
 \ifnum #1\expandafter \@firstoftwo
 \else \expandafter \@secondoftwo
 \fi
}%
\providecommand \@ifx [1]{%
 \ifx #1\expandafter \@firstoftwo
 \else \expandafter \@secondoftwo
 \fi
}%
\providecommand \natexlab [1]{#1}%
\providecommand \enquote  [1]{``#1''}%
\providecommand \bibnamefont  [1]{#1}%
\providecommand \bibfnamefont [1]{#1}%
\providecommand \citenamefont [1]{#1}%
\providecommand \href@noop [0]{\@secondoftwo}%
\providecommand \href [0]{\begingroup \@sanitize@url \@href}%
\providecommand \@href[1]{\@@startlink{#1}\@@href}%
\providecommand \@@href[1]{\endgroup#1\@@endlink}%
\providecommand \@sanitize@url [0]{\catcode `\\12\catcode `\$12\catcode
  `\&12\catcode `\#12\catcode `\^12\catcode `\_12\catcode `\%12\relax}%
\providecommand \@@startlink[1]{}%
\providecommand \@@endlink[0]{}%
\providecommand \url  [0]{\begingroup\@sanitize@url \@url }%
\providecommand \@url [1]{\endgroup\@href {#1}{\urlprefix }}%
\providecommand \urlprefix  [0]{URL }%
\providecommand \Eprint [0]{\href }%
\providecommand \doibase [0]{http://dx.doi.org/}%
\providecommand \selectlanguage [0]{\@gobble}%
\providecommand \bibinfo  [0]{\@secondoftwo}%
\providecommand \bibfield  [0]{\@secondoftwo}%
\providecommand \translation [1]{[#1]}%
\providecommand \BibitemOpen [0]{}%
\providecommand \bibitemStop [0]{}%
\providecommand \bibitemNoStop [0]{.\EOS\space}%
\providecommand \EOS [0]{\spacefactor3000\relax}%
\providecommand \BibitemShut  [1]{\csname bibitem#1\endcsname}%
\let\auto@bib@innerbib\@empty
\bibitem [{\citenamefont {Horodecki}\ \emph {et~al.}(2005)\citenamefont
  {Horodecki}, \citenamefont {Horodecki}, \citenamefont {Horodecki},\ and\
  \citenamefont {Oppenheim}}]{pptkey}%
  \BibitemOpen
  \bibfield  {author} {\bibinfo {author} {\bibfnamefont {K.}~\bibnamefont
  {Horodecki}}, \bibinfo {author} {\bibfnamefont {M.}~\bibnamefont
  {Horodecki}}, \bibinfo {author} {\bibfnamefont {P.}~\bibnamefont
  {Horodecki}}, \ and\ \bibinfo {author} {\bibfnamefont {J.}~\bibnamefont
  {Oppenheim}},\ }\href@noop {} {\bibfield  {journal} {\bibinfo  {journal}
  {Phys. Rev. Lett.}\ }\textbf {\bibinfo {volume} {94}},\ \bibinfo {pages}
  {160502} (\bibinfo {year} {2005})},\ \Eprint
  {http://arxiv.org/abs/quant-ph/0309110} {quant-ph/0309110} \BibitemShut
  {NoStop}%
\bibitem [{\citenamefont {Horodecki}\ \emph {et~al.}(2006)\citenamefont
  {Horodecki}, \citenamefont {Horodecki}, \citenamefont {Horodecki},
  \citenamefont {Leung},\ and\ \citenamefont
  {Oppenheim}}]{horodecki_qkdprivst}%
  \BibitemOpen
  \bibfield  {author} {\bibinfo {author} {\bibfnamefont {K.}~\bibnamefont
  {Horodecki}}, \bibinfo {author} {\bibfnamefont {M.}~\bibnamefont
  {Horodecki}}, \bibinfo {author} {\bibfnamefont {P.}~\bibnamefont
  {Horodecki}}, \bibinfo {author} {\bibfnamefont {D.}~\bibnamefont {Leung}}, \
  and\ \bibinfo {author} {\bibfnamefont {J.}~\bibnamefont {Oppenheim}},\ }\href
  {http://arxiv.org/abs/quant-ph/0608195} {\bibfield  {journal} {\bibinfo
  {journal} {quant-ph/0608195}\ } (\bibinfo {year} {2006})},\ \bibinfo {note}
  {{IEEE} Transactions on Information Theory Vol 54 Issue 6 p2604-2620
  (2008)}\BibitemShut {NoStop}%
\bibitem[{\citenamefont{Smith and Yard}(2008)}]{smith_superact}
\bibinfo{author}{\bibfnamefont{G.}~\bibnamefont{Smith}} \bibnamefont{and}
  \bibinfo{author}{\bibfnamefont{J.}~\bibnamefont{Yard}},
  \bibinfo{journal}{Science} \textbf{\bibinfo{volume}{321}},
  \bibinfo{pages}{1812 } (\bibinfo{year}{2008})
\bibitem [{\citenamefont {Smith}\ and\ \citenamefont
  {Smolin}(2009)}]{smith_extensive_2009}%
  \BibitemOpen
  \bibfield  {author} {\bibinfo {author} {\bibfnamefont {G.}~\bibnamefont
  {Smith}}\ and\ \bibinfo {author} {\bibfnamefont {J.~A.}\ \bibnamefont
  {Smolin}},\ }\href {\doibase 10.1103/PhysRevLett.103.120503} {\bibfield
  {journal} {\bibinfo  {journal} {Physical Review Letters}\ }\textbf {\bibinfo
  {volume} {103}},\ \bibinfo {pages} {120503} (\bibinfo {year}
  {2009})}\BibitemShut {NoStop}%
\bibitem[{\citenamefont{Cubitt et~al.}(2008)\citenamefont{Cubitt, Ruskai, and
  Smith}}]{cubitt_structuredqc}
\bibinfo{author}{\bibfnamefont{T.~S.} \bibnamefont{Cubitt}},
  \bibinfo{author}{\bibfnamefont{M.~B.} \bibnamefont{Ruskai}},
  \bibnamefont{and} \bibinfo{author}{\bibfnamefont{G.}~\bibnamefont{Smith}},
  \bibinfo{journal}{Journal of Mathematical Physics}
  \textbf{\bibinfo{volume}{49}}, \bibinfo{pages}{102104}
  (\bibinfo{year}{2008})
\bibitem [{\citenamefont {Peres}(1996)}]{peres_separabilitycr}%
  \BibitemOpen
  \bibfield  {author} {\bibinfo {author} {\bibfnamefont {A.}~\bibnamefont
  {Peres}},\ }\href {\doibase 10.1103/PhysRevLett.77.1413} {\bibfield
  {journal} {\bibinfo  {journal} {Physical Review Letters}\ }\textbf {\bibinfo
  {volume} {77}},\ \bibinfo {pages} {1413} (\bibinfo {year}
  {1996})}\BibitemShut {NoStop}%
\bibitem [{\citenamefont {Horodecki}\ \emph {et~al.}(1999)\citenamefont
  {Horodecki}, \citenamefont {Horodecki},\ and\ \citenamefont
  {Horodecki}}]{horodecki_bec}%
  \BibitemOpen
  \bibfield  {author} {\bibinfo {author} {\bibfnamefont {P.}~\bibnamefont
  {Horodecki}}, \bibinfo {author} {\bibfnamefont {M.}~\bibnamefont
  {Horodecki}}, \ and\ \bibinfo {author} {\bibfnamefont {R.}~\bibnamefont
  {Horodecki}},\ }\href {http://arxiv.org/abs/quant-ph/9904092} {\bibfield
  {journal} {\bibinfo  {journal} {quant-ph/9904092}\ } (\bibinfo {year}
  {1999})},\ \bibinfo {note} {{J.Mod.Opt.} 47 (2000) 347-354}\BibitemShut
  {NoStop}%
\bibitem [{\citenamefont {Bennett}\ \emph {et~al.}(1996)\citenamefont
  {Bennett}, \citenamefont {DiVincenzo}, \citenamefont {Smolin},\ and\
  \citenamefont {Wootters}}]{depolarizing}%
  \BibitemOpen
  \bibfield  {author} {\bibinfo {author} {\bibfnamefont {C.~H.}\ \bibnamefont
  {Bennett}}, \bibinfo {author} {\bibfnamefont {D.~P.}\ \bibnamefont
  {DiVincenzo}}, \bibinfo {author} {\bibfnamefont {J.~A.}\ \bibnamefont
  {Smolin}}, \ and\ \bibinfo {author} {\bibfnamefont {W.~K.}\ \bibnamefont
  {Wootters}},\ }\href {\doibase 10.1103/PhysRevA.54.3824} {\bibfield
  {journal} {\bibinfo  {journal} {Phys. Rev. A}\ }\textbf {\bibinfo {volume}
  {54}},\ \bibinfo {pages} {3824} (\bibinfo {year} {1996})}\BibitemShut
  {NoStop}%
\bibitem[{\citenamefont{Smith et~al.}(2011)\citenamefont{Smith, Smolin, and
  Yard}}]{smith_gaussian_2011}
\bibinfo{author}{\bibfnamefont{G.}~\bibnamefont{Smith}},
  \bibinfo{author}{\bibfnamefont{J.~A.} \bibnamefont{Smolin}},
  \bibnamefont{and} \bibinfo{author}{\bibfnamefont{J.}~\bibnamefont{Yard}},
  \bibinfo{journal}{arXiv:1102.4580}  (\bibinfo{year}{2011}).

\bibitem [{\citenamefont {Lloyd}(1997)}]{Lloyd-cap}%
  \BibitemOpen
  \bibfield  {author} {\bibinfo {author} {\bibfnamefont {S.}~\bibnamefont
  {Lloyd}},\ }\href@noop {} {\bibfield  {journal} {\bibinfo  {journal} {Phys.
  Rev. A}\ }\textbf {\bibinfo {volume} {55}},\ \bibinfo {pages} {1613}
  (\bibinfo {year} {1997})},\ \Eprint {http://arxiv.org/abs/quant-ph/9604015}
  {quant-ph/9604015} \BibitemShut {NoStop}%
\bibitem [{\citenamefont {Shor}(2002)}]{shor-cap}%
  \BibitemOpen
  \bibfield  {author} {\bibinfo {author} {\bibfnamefont {P.}~\bibnamefont
  {Shor}},\ }\href@noop {} {\enquote {\bibinfo {title} {The quantum channel
  capacity and coherent information, MSRI workshop on quantum computation},}\
  } (\bibinfo {year} {2002}),\ \bibinfo {note} {(Avaliable at
  http://www.msri.org/publications/ln/msri/2002/quantumcrypto/shor/1/)}\BibitemShut
  {NoStop}%
\bibitem [{\citenamefont {Devetak}(2003)}]{devetak_privatecc}%
  \BibitemOpen
  \bibfield  {author} {\bibinfo {author} {\bibfnamefont {I.}~\bibnamefont
  {Devetak}},\ }\href {http://arxiv.org/abs/quant-ph/0304127} {\bibfield
  {journal} {\bibinfo  {journal} {quant-ph/0304127}\ } (\bibinfo {year}
  {2003})}\BibitemShut {NoStop}%
\bibitem [{\citenamefont {Horodecki}\ \emph {et~al.}(1998)\citenamefont
  {Horodecki}, \citenamefont {Horodecki},\ and\ \citenamefont
  {Horodecki}}]{horodecki_BE}%
  \BibitemOpen
  \bibfield  {author} {\bibinfo {author} {\bibfnamefont {M.}~\bibnamefont
  {Horodecki}}, \bibinfo {author} {\bibfnamefont {P.}~\bibnamefont
  {Horodecki}}, \ and\ \bibinfo {author} {\bibfnamefont {R.}~\bibnamefont
  {Horodecki}},\ }\href {\doibase 10.1103/PhysRevLett.80.5239} {\bibfield
  {journal} {\bibinfo  {journal} {Physical Review Letters}\ }\textbf {\bibinfo
  {volume} {80}},\ \bibinfo {pages} {5239} (\bibinfo {year}
  {1998})}\BibitemShut {NoStop}%
\bibitem [{\citenamefont {Horodecki}\ \emph {et~al.}(2009)\citenamefont
  {Horodecki}, \citenamefont {Horodecki}, \citenamefont {Horodecki},\ and\
  \citenamefont {Oppenheim}}]{horodecki_general_2009}%
  \BibitemOpen
  \bibfield  {author} {\bibinfo {author} {\bibfnamefont {K.}~\bibnamefont
  {Horodecki}}, \bibinfo {author} {\bibfnamefont {M.}~\bibnamefont
  {Horodecki}}, \bibinfo {author} {\bibfnamefont {P.}~\bibnamefont
  {Horodecki}}, \ and\ \bibinfo {author} {\bibfnamefont {J.}~\bibnamefont
  {Oppenheim}},\ }\href {\doibase 10.1109/TIT.2008.2009798} {\bibfield
  {journal} {\bibinfo  {journal} {{IEEE} Transactions on Information Theory}\
  }\textbf {\bibinfo {volume} {55}},\ \bibinfo {pages} {1898} (\bibinfo {year}
  {2009})}\BibitemShut {NoStop}%
\bibitem [{\citenamefont {Oppenheim}(2008)}]{O08-superactivation}%
  \BibitemOpen
  \bibfield  {author} {\bibinfo {author} {\bibfnamefont {J.}~\bibnamefont
  {Oppenheim}},\ }\href@noop {} {\bibfield  {journal} {\bibinfo  {journal}
  {Science}\ }\textbf {\bibinfo {volume} {321}},\ \bibinfo {pages} {1783}
  (\bibinfo {year} {2008})}\BibitemShut {NoStop}%
\bibitem [{\citenamefont {Hastings}()}]{hastings_infinitely_many_chn}%
  \BibitemOpen
  \bibfield  {author} {\bibinfo {author} {\bibfnamefont {M.}~\bibnamefont
  {Hastings}},\ }\href
  {http://permalink.lanl.gov/object/tr?what=info:lanl-repo/lareport/LA-UR-08-07577}
  {\bibinfo  {journal} {{LANL;} {RN10098910, and M. Hastings (private communication)}}\ }\BibitemShut {NoStop}%
\bibitem [{\citenamefont {Alicki}\ and\ \citenamefont
  {Fannes}(2004)}]{alicki_fannes}%
  \BibitemOpen
\bibfield  {journal} {  }\bibfield  {author} {\bibinfo {author} {\bibfnamefont
  {R.}~\bibnamefont {Alicki}}\ and\ \bibinfo {author} {\bibfnamefont
  {M.}~\bibnamefont {Fannes}},\ }\href@noop {} {\bibfield  {journal} {\bibinfo
  {journal} {Journal of Physics A}\ }\textbf {\bibinfo {volume} {37}} (\bibinfo
  {year} {2004})}\BibitemShut {NoStop}%
 \bibitem [{\citenamefont {Wootters}\ and\ \citenamefont
  {Zurek}(1982)}]{WoottersZ-cloning}%
  \BibitemOpen
  \bibfield  {author} {\bibinfo {author} {\bibfnamefont {W.~K.}\ \bibnamefont
  {Wootters}}\ and\ \bibinfo {author} {\bibfnamefont {W.~H.}\ \bibnamefont
  {Zurek}},\ }\href@noop {} {\bibfield  {journal} {\bibinfo  {journal}
  {Nature}\ }\textbf {\bibinfo {volume} {299}},\ \bibinfo {pages} {802}
  (\bibinfo {year} {1982})}\BibitemShut {NoStop}%
\end{thebibliography}
\end{document}